\documentclass[11pt,aps,prd,amsmath,amssymb,showpacs=false,floats,floatfix,nofootinbib,superscriptaddress]{revtex4}
\usepackage{graphicx} % Required for inserting images
\usepackage[colorlinks=true, linkcolor=red, citecolor=blue,CJKbookmarks=true]{hyperref}
\usepackage{amsmath,amsfonts,amssymb}
\usepackage{float}
\usepackage{graphicx}
\usepackage{color}
\usepackage[dvipsnames,svgnames,x11names]{xcolor}
\usepackage{slashed} % slash mark
\usepackage{url}
\usepackage{subfigure}
\usepackage{multirow} % multirows in tabular environment
\usepackage{unicode-math}
\usepackage{xeCJK}
\graphicspath{{figs/}}  % figure path
\usepackage[
  markup=default,           % 高亮风格：default / underlined / highlighted
  authormarkup=none,        % 隐藏作者 id 上标
]{changes}
\newsavebox{\zmstrikebox}
\newcommand{\mstrike}[2][0.5pt]{%
  \sbox{\zmstrikebox}{$#2$}%
  \makebox[0pt][l]{\usebox{\zmstrikebox}}%
  \rule[0.5\ht\zmstrikebox-#1/2]{\wd\zmstrikebox}{#1}}
\setdeletedmarkup{\ifmmode\textcolor{CornflowerBlue}{\mstrike{#1}}\else\textcolor{CornflowerBlue}{\sout{#1}}\fi}
\usepackage{letltxmacro}
\LetLtxMacro{\savedreplaced}{\replaced}
\makeatletter
\renewcommand{\replaced}[3][\@empty]{%
  \savedreplaced[#1]{#2\ }{#3}%
}
\makeatother
\let\oldcite\cite
\renewcommand{\cite}[1]{\mbox{\oldcite{#1}}}
\usepackage{amssymb,amsmath}

\def\GeV{\mathrm{GeV}} % GeV
\def\hate{\hat{\mathbf{e}}}

\definechangesauthor[color=OrangeRed]{zhm}
\definechangesauthor[color=Pink]{jfeng}
\definechangesauthor[color=ForestGreen]{lsj}

\begin{document}
%\linenumbers
\title{A Unified Charge-Dependent Modulation Model for AMS-02 Proton and Antiproton Fluxes during Solar Minimum}
\author{Hui-Ming Zhang}
\affiliation{School of Physics and Astronomy, Sun Yat-sen University, Zhuhai, 519082, China}

\author{Su-Jie Lin}
\email{linsj@ouc.edu.cn}
\affiliation{State Key Laboratory of Physical Oceanography, Ocean University of China, Qingdao 266100, China}
\affiliation{School of Physics and Astronomy, Sun Yat-sen University, Zhuhai, 519082, China}

\author{Jie Feng}
\email{fengj77@mail.sysu.edu.cn}
\affiliation{School of Science, Shenzhen Campus of Sun Yat-sen University, Shenzhen, 518107, China}
\affiliation{Guangdong Provincial Key Laboratory of Advanced Particle Detection Technology, Dongguan, 523803, China}

\author{Jie-Teng Jiang}
\affiliation{Institute of Science and Technology for Deep Space Exploration, Suzhou Campus, Nanjing University, Suzhou, 215163, China}

\author{Yu-Dong Cui}
\affiliation{School of Physics and Astronomy, Sun Yat-sen University, Zhuhai, 519082, China}

\author{Yi-Han Liu}
\affiliation{School of Physics and Astronomy, Sun Yat-sen University, Zhuhai, 519082, China}

\author{Lili Yang}
\email{yanglli5@mail.sysu.edu.cn}
\affiliation{School of Physics and Astronomy, Sun Yat-sen University, Zhuhai, 519082, China}

\pacs{96.50.sb, 96.60.-j}
 
\begin{abstract}
  We develop a unified charge-dependent solar modulation model by solving the three-dimensional Parker transport equation, incorporating a realistic wavy heliospheric current sheet to treat drift effects self-consistently. Using a local interstellar spectrum from GALPROP constrained by Voyager data, we fit the model to time-resolved proton and antiproton fluxes measured by the Alpha Magnetic Spectrometer-02 (AMS-02) from May 2011 to June 2022, excluding the heliospheric magnetic field (HMF) polarity-reversal epoch (mid-2012 to early 2015). To enable rapid parameter scans, we employ neural-network-based surrogate models to compute propagation and modulation matrices efficiently. The results demonstrate that the model simultaneously describes the observed proton and antiproton fluxes with physically reasonable parameters, providing a unified account of charge-dependent modulation.
\end{abstract}
\maketitle

\section{Introduction}

Galactic cosmic rays (GCRs) detected at the top of the atmosphere (TOA) originate from sources such as supernova remnants (SNRs).
After propagating through the galaxy, they undergo modulation by the heliospheric magnetic field (HMF) upon entering the solar system.
The injection, propagation and modulation processes collectively shape the GCR spectrum observed at TOA.
Consequently, the interdependence of these processes complicates the interpretation of cosmic ray flux, especially in the GeV energy range, obscuring potential anomalous signals, such as the low-energy secondary particles from dark matter~\cite{linInvestigatingDarkMatter2019,cuiPossibleDarkMatter2017,yuanPropagationCosmicRays2017,zhuExplainingGeVAntiproton2022}.

Two strategies effectively disentangle the modulation process from the injection and propagation processes: examining the local interstellar spectrum (LIS) and studying temporal variations in the cosmic-ray (CR) spectrum.
The CR LIS, entirely free of heliospheric modulation, solely reflects injection and propagation effects.
However, only the two Voyager spacecrafts have traversed beyond the heliosphere to date~\cite{cummingsGALACTICCOSMICRAYS2016,stoneCosmicRayMeasurements2019}. Their CR observations, spanning MeV to sub-GeV energies, remain the only available LIS measurements.
Alternatively, temporal variations in the TOA spectrum can separate modulation effects, as the HMF varies over the 11-year solar cycle while injection and propagation processes remain constant.
The AMS-02, Payload for Antimatter Matter Exploration and Light-nuclei Astrophysics (PAMELA), and High Energy Cosmic-radiation Detection (HERD) experiments have provided precise, time-resolved CR flux measurements over a full solar cycle~\cite{adrianiTimeDependenceProton2013,amscollaborationTemporalStructuresElectron2023,amscollaborationTemporalStructuresPositron2023,aguilarPeriodicitiesDailyProton2021,amscollaborationAntiprotonsElementaryParticles2025,xuPotentialConstrainingPropagation2023}.
Combined with Voyager's sub-GeV LIS data, these measurements can be employed to illuminate modulation effects.

Upon encountering the heliosphere, cosmic ray (CR) particles diffuse through the HMF.
The structure of HMF induces charge-dependent drifts, while the solar wind drives convection and adiabatic energy losses.
Particles of opposite charges drift in opposing directions, resulting in distinct diffusion paths and modulation strengths.
Phenomenologically, the force field approximation (FFA) describes this effect by assigning distinct modulation potentials to each charge sign~\cite{gleesonSolarModulationGalactic1968, cholisPredictiveAnalyticModel2016, kuhlenTimeDependentAMS02ElectronPositron2019,lipariSpectralShapesFluxes2019, silverTestingCosmicRayPropagation2024, koldobskiyValidationNeutronMonitor, wangTimeDependentSolar2019}.
Within this framework, particles sharing the same charge sign have been found to exhibit consistent modulation behavior~\cite{zhaoUnifiedSolarModulation2025}.
Although the FFA with charge-dependent modulation potential phenomenologically characterizes drifts, it neither elucidates the underlying drift mechanisms nor predicts modulation differences induced by drift.

Beyond the phenomenological FFA model, more sophisticated models based on the time-backward stochastic differential equation (SDE) solution of Parker's transport equation have been developed to simultaneously describe modulation effects on particles of different charge signs.
These models, whether in two dimensions \cite{kapplSOLARPROPChargesignDependent2016,jiangNewScenarioSolar2023} or three \cite{potgieterModulationGalacticProtons2014,songNumericalStudySolar2021,duanScrutinizingImpactSolar2025}, can simulate both positively and negatively charged particles under identical physical conditions.
Among these, the HelMod model treats the transport of protons, antiprotons, electrons, positrons, and nuclei within a single framework~\cite{boschiniPropagationCosmicRays2018,bobikAntiprotonModulationHeliosphere2011,dellaTorreEffectsSolarModulation2012}, and its predictions have been extensively compared with AMS-02 measurements.
Using the time-resolved fluxes of protons, antiprotons, electrons, and positrons measured by the AMS-02 collaboration from May 2011 to June 2022\cite{amscollaborationAntiprotonsElementaryParticles2025}, we investigate drift effects with such models, and adopt a recent 3D heliospheric current sheet (HCS) drift model derived from a realistic wavy HCS structure\cite{peiThreedimensionalWavyHeliospheric2012}.
By incorporating the LIS obtained from the GALPROP package~\cite{moskalenkoProductionPropagationCosmicRay1998,strongPropagationCosmicRayNucleons1998a}, we demonstrate that this model has potential to provide a unified description of modulation for all four elementary cosmic-ray species, irrespective of charge or mass, and is fitted to Voyager and AMS-02 measurements.
Such a unified framework can be further applied to study potential anomalous signals obscured in the GeV-range cosmic-ray spectra.

% \added[id=zhm]{
% Leveraging recently published time-resolved fluxes of protons, antiprotons, electrons, and positrons from the AMS-02 collaboration (May 2011–June 2022)~\cite{amscollaborationAntiprotonsElementaryParticles2025}, we perform a refined analysis of CR drift effects in the heliosphere.
% Extending conventional approaches to heliospheric current sheet (HCS) drift parameterization~\cite{duanScrutinizingImpactSolar2025}, we numerically solve the three-dimensional Parker transport equation for CR diffusion in the HMF using a time-backward stochastic method~\cite{jiangNewScenarioSolar2023}, aligning with the SolarProp package methodology~\cite{kapplSOLARPROPChargesignDependent2016}.
% Besides, A key advancement lies in explicitly modeling a realistic wavy HCS structure to characterize HCS drift, a component distinct from curvature and gradient drifts in the HMF~\cite{peiThreedimensionalWavyHeliospheric2012}.
% By implementing the LIS derived from the GALPROP package~\cite{moskalenkoProductionPropagationCosmicRay1998,strongPropagationCosmicRayNucleons1998a}, this model achieves a unified description of CR modulation across all elementary particle species, demonstrating consistent agreement with both Voyager and AMS-02 observations. The methodology further enables identification of anomalous spectral features previously obscured in GeV-range CR measurements.
% }

This paper is structured as follows. Section~\ref{sec:modulation_model} introduces the Parker transport equation, including the structure of the heliospheric magnetic field (HMF) and heliospheric current sheet (HCS), along with its stochastic solution.
Section~\ref{sec:results} presents the model parameter settings and fitting results.
Section~\ref{sec:conclusion} provides the discussion and conclusions.

\section{Local Interstellar Spectra}

After being accelerated at their sources, CR particles propagate through the Galactic magnetic field, undergoing energy losses, fragmentation, convection, and reacceleration processes.
These particles eventually reach the local interstellar region before encountering the Solar System.
The LIS within this region serves as the fundamental input that determines the modulated spectra observed at the TOA.

The LIS has been measured by the two Voyager spacecraft directly, which remain the only human-made objects that have crossed the heliopause and exited the heliosphere.
Their Cosmic Ray Subsystem (CRS) detects electrons/positrons in the energy range of 3-110 MeV and nuclei from 1-500 MeV/nucleon.
However, as this system is unable to discriminate particle charge signs, Voyager data cannot determine the LIS of secondary species such as antiprotons and positrons, despite providing crucial direct measurements of primary cosmic rays like protons.

Direct measurements of local interstellar cosmic rays (CRs) are unavailable for energies above 0.5 GeV, necessitating a model to represent the local interstellar spectrum (LIS) in this range.
For primary species such as protons and electrons, a spline interpolation model spanning MeV to tens of GeV can be applied to describe the LIS.
In this work, however, we aim to develop a unified description for not only these primary particles but also secondary species, specifically antiprotons and positrons.
To achieve a physically consistent description of all these species, we utilize the GALPROP package, following the same philosophy as the self-consistent GALPROP--HelMod framework in which the parameters of the two models are cross-calibrated~\cite{boschiniSolutionHeliosphericPropagation2017,boschiniInferenceLocalInterstellar2020}.

In the GALPROP model, the injection of CRs into the Galaxy is described by a triple broken power-law spectrum, defined by spectral indices \(-\nu_1\), \(-\nu_2\), and \(-\nu_3\), along with break rigidities \(R_\mathrm{br,1}\) and \(R_\mathrm{br,2}\).
After injection, CRs propagate through the Galactic halo of half-height \(z_h\), undergoing diffusion and reacceleration before reaching the local interstellar medium.
This reacceleration is characterized by the Alfv\'{e}n speed \(v_A\).
The diffusion coefficient is governed by the particle speed \(\beta\) (in natural units) and rigidity \(R\), in relation
\begin{equation}
  D = D_0\beta^\eta
  \begin{cases}
    \left(\dfrac{R}{R_0}\right)^\delta, & R < R_h, \\
    \left(\dfrac{R_h}{R_0}\right)^\delta\left(\dfrac{R}{R_h}\right)^{\delta_h}, & R \ge R_h,
  \end{cases}
  \label{eq:D_coefficient}
\end{equation}
where the reference rigidity \(R_0\) is fixed at \(4\ \mathrm{GV}\).
The broken-power-law form is introduced primarily to accommodate the high-energy boron-to-carbon (B/C) ratio measurements from AMS-02 and DAMPE~\cite{genoliniCosmicrayTransportAMS022019, maInterpretationsCosmicRay2023}. In the present analysis, however, we do not consider data in that high-energy regime. We therefore exclude B/C measurements above 200 GeV and, correspondingly, omit the high-energy break parameters $R_h$ and $\delta_h$ in the following analysis.

For this work, we adopt the ranges of these injection and propagation parameters from the recent study~\cite{maInterpretationsCosmicRay2023}, which constrained them using DAMPE measurements.
The corresponding parameter ranges are listed in Table~\ref{tab:propagation_parameters}.

\begin{table}[h]
    \centering
    \begin{tabular}{lll}
        \hline
        \textbf{Parameter} & \textbf{Range} & \textbf{Units} \\
        \hline
        $\nu_1$           & [1.5  , 2.5  ] & - \\
        $\nu_2$           & [2.2  , 2.5  ] & - \\
        $\nu_3$           & [1.5  , 3.0  ] & - \\
        $R_\mathrm{br,1}$ & [1    , 80   ] & GV \\
        $R_\mathrm{br,2}$ & [100  , 10000] & GV \\
        \hline
        $D_0$             & [1    , 20   ] & $\mathrm{10^{28} cm^2\,s^{-1}}$ \\
        $\eta$             & [-3   , 1.5  ] & - \\
        $\delta$           & [0.3  , 0.6  ] & - \\
        $v_A$             & [6    , 80   ] & $\mathrm{km\,s^{-1}}$ \\
        $z_h$             & [1.5  , 15   ] & kpc \\
        \hline
    \end{tabular}
    \caption{Allowed ranges for Galactic CR injection and propagation parameters.}
    \label{tab:propagation_parameters}
\end{table}

\section{Modulation Model} \label{sec:modulation_model}

The CRs entering the heliosphere encounter the heliospheric magnetic field (HMF), which is embedded within the solar wind. Their transport is governed by Parker's equation~\cite{parkerPassageEnergeticCharged1965}:
\begin{equation}
\begin{aligned}
    \frac{\partial f(\mathbf{r},\mathbf{p},t)}{\partial t} = &\nabla\cdot\left(\mathbf{K}\cdot\nabla f(\mathbf{r},\mathbf{p},t) - \mathbf{V_{SW}}f(\mathbf{r},\mathbf{p}, t)\right) \\
    & + \frac{1}{3}(\nabla\cdot \mathbf{V_{SW}})\frac{\partial f(\mathbf{r},\mathbf{p},t)}{\partial \ln\,p}
    \label{eq:TPE}
\end{aligned}
\end{equation}
where $f(\mathbf{r},\mathbf{p},t)$ is the phase-space distribution function of galactic cosmic rays (GCRs).
The three terms on the right-hand side of Equation~\eqref{eq:TPE} describe diffusion, adiabatic energy loss, and convection, respectively.
The propagation of CRs in the HMF is governed by the tensor $\mathbf{K}$.
The solar wind velocity, $\mathbf{V_{SW}}$, drives both the adiabatic energy loss and convection processes.
We then describe the assumed solar wind velocity ($\mathbf{V_{SW}}$) distribution and corresponding HMF structure, along with the detailed diffusion model in the following subsections, to outline the implementation of our modulation model.

\subsection{Heliosphere Structure}

The solar wind, emanating from the Sun, extends magnetic field lines into interplanetary space, forming the HMF.
Frozen into the solar wind plasma and rooted in the Sun's northern and southern hemispheres, these field lines establish two distinct domains of opposite magnetic polarity---outward and inward, respectively.
Their interface forms the heliospheric current sheet (HCS), a thin layer of electric current induced by the interaction of these oppositely directed magnetic fields.
Due to the Sun's rotation together with the tilt of its magnetic axis, both the HMF and HCS are drawn into spiral configurations: the field lines form the archetypal Parker's spiral~\cite{parkerDynamicsInterplanetaryGas1958}, while the HCS is warped into a complex, wavy structure resembling a ballerina's skirt.

The large-scale structure of the spiral HMF can be described by
\begin{equation}
\mathbf{B} = \frac{B_0r_0^2}{r^2}\left(\hate_r - \frac{\Omega_\odot r}{\mathrm{V_{SW}}}\sin\theta\hate_\phi\right)\left[1 - 2 \mathrm{H}(\theta - \theta_\mathrm{cs})\right],
\label{eq:B_field}
\end{equation}
where $B_0$ is the reference magnetic field strength at a distance $r_0$, $\Omega_\odot$ is the solar angular rotation velocity, which we take to correspond to a period of $27.5$ days, and $\mathrm{H}(\theta-\theta_\mathrm{cs})$ is the Heaviside step function with $\theta_\mathrm{cs}$ representing the polar angle of the HCS. This period is close to the synodic Carrington rotation period rather than to the sidereal rotation period of the Sun~\cite{snodgrassRotationDopplerFeatures1990}. We have verified that adopting instead the sidereal equatorial value of $24.5$ days changes the modelled proton and antiproton spectra at $1\,\mathrm{AU}$ by at most $1$\,\%, so that this choice does not affect any of the conclusions drawn in this work.
Gauss's law for magnetism necessitates the conservation of the quantity $B_0 r_0^2$ throughout the HMF.
Transverse perturbations of the HMF near the Sun can significantly enhance the field magnitude in polar regions, modifying the archetypal spiral and introducing a $\hate_\theta$ component~\cite{fichtnerCosmicRayModulation1996,jokipiiPolarHeliosphericMagnetic1989}.
Such a polar correction is not negligible in general: it modifies the latitudinal gradients of the GCR intensity and, in certain HMF configurations, particles enter the heliosphere predominantly through the polar regions~\cite{bobikSystematicInvestigationSolar2012,bobikLatitudinalDependenceCosmic2013}. Since only limited measurements of the polar HMF from Ulysses~\cite{baloghHeliosphericMagneticField1995} are available, and since the polar correction cannot be constrained by the data employed here, we adopt the standard Parker spiral model.
Future observations from the second satellite of the Kuafu project~\cite{tuIntroductionKuaFuProject2006} may warrant implementation of an updated HMF model.

The polar angle of the HCS $\theta_\mathrm{cs}$ varies with both radial distance $r$ and azimuthal angle $\phi$.
This spatial dependence determines the HCS shape through the relation
\begin{equation}
\tan\left(\frac{\pi}{2} - \theta_\mathrm{cs}\right) = \tan\alpha\sin\left[\phi + \frac{\Omega_\odot r}{\mathrm{V_{SW}}} - \Omega_\odot(t - t_0)\right],
\label{eq:theta_cs}
\end{equation}
where $\alpha$ is the tilt angle between the Sun's magnetic and rotation axes.
While this equation was originally approximated using a sine function for small tilt angles~\cite{jokipiiEffectsDriftTransport1981}, the exact tangent form of the wavy current sheet adopted here follows Refs.~\cite{kotaEffectsDriftTransport1983,thomasEffectHeliosphericCurrent1986,burgerInclusionWavyNeutral1987,peiThreedimensionalWavyHeliospheric2012}.
Although $\alpha$ varies in time and space under realistic conditions, we adopt a constant value for simplicity in this work.

The solar wind velocity is radially aligned and has been observed to vary with heliocentric distance $r$ and colatitude $\theta$ during solar minimum~\cite{bameUlyssesSolarWind1992, heberCosmicRaysHigh2006}.
Inside the termination shock (TS), the wind speed $\mathrm{V_{SW}}$ remains constant with heliocentric distance, ranging from approximately $430\,\mathrm{km\,s^{-1}}$ at the equatorial plane to about $750\,\mathrm{km\,s^{-1}}$ in the polar regions~\cite{mccomasWeakerSolarWind2008}. Beyond the TS, the wind speed decreases sharply, dropping to roughly $170\,\mathrm{km\,s^{-1}}$ at the equatorial plane~\cite{potgieterModulationGalacticProtons2014}.
For the region extending from approximately $1\,\mathrm{AU}$ to the TS, we adopt the solar wind speed profile from Ref.~\cite{potgieterModulationGalacticProtons2014}:
\begin{equation}
\begin{aligned}
\mathbf{V_{SW}}(r, \theta) =& V_0\left(1.475\mp 0.4\tanh\left[6.8\left(\theta - \frac{\pi}{2} \pm \theta_0\right)\right]\right) \\
&\times \left(\frac{s + 1}{2s} - \frac{s - 1}{2s}\tanh\frac{r - r_\mathrm{TS}}{L}\right)\hate_r,
\label{eq:SW_speed}
\end{aligned}
\end{equation}
where $V_0 = 400\,\mathrm{km\,s^{-1}}$, $s = 2.5$, $L = 1.2\,\mathrm{AU}$, and the TS distance $r_\mathrm{TS} = 90\,\mathrm{AU}$~\cite{potgieterModulationGalacticProtons2014}.
The top and bottom signs correspond to the northern and southern hemispheres, respectively, with $\theta_0 \equiv \alpha + 15\pi/180$ defining the boundary between the slow and fast solar wind regions.
Equation~\eqref{eq:SW_speed} yields wind speeds consistent with observations: approximately $430\,\mathrm{km\,s^{-1}}$ in the slow region near the equator and about $750\,\mathrm{km\,s^{-1}}$ in the fast region near the poles.
These values apply to solar minimum conditions; in Eq.~\eqref{eq:SW_speed}, the latitudinal contrast weakens as activity increases, because the boundary $\theta_0=\alpha+15^\circ$ shifts towards higher latitudes as $\alpha$ grows. During the unusually weak minimum of 2006--2009, the polar coronal holes produced a substantially slower wind~\cite{mccomasWeakerSolarWind2008}.

\subsection{Diffusion Model}

The propagation of CR particles through the HMF includes diffusion and drift. The diffusion, driven by the turbulent field component, is governed by the symmetric part of the diffusion tensor, $\mathbf{K}_S$.
Conversely, the drift, resulting from the regular HMF component, can be described by the antisymmetric part, $\mathbf{K}_A$.
Defined with respect to the direction of the HMF, the full diffusion tensor $\mathbf{K}$ is given by:
\begin{equation}
  \begin{aligned}
    \mathbf{K} &= \mathbf{K}_S + \mathbf{K}_A \\
    &= \begin{pmatrix}
        K_{\perp r} && \\
        & K_{\perp\theta} &\\
        && K_\parallel
       \end{pmatrix} +
       \begin{pmatrix}
       & - K_A & \\
       K_A && \\
       && 0
       \end{pmatrix}.
  \end{aligned}
  \label{eq:tensor_K}
\end{equation}

In the symmetric component of the diffusion tensor, $K_\parallel$ is the diffusion coefficient parallel to the magnetic field, while $K_{\perp r}$ and $K_{\perp\theta}$ are the two perpendicular coefficients, whose directions are the projections of the radial and colatitude directions onto the plane perpendicular to $\mathbf{B}$; the latter projection is exact because the Parker field of Eq.~\eqref{eq:B_field} has no $\hate_\theta$ component. Expressing the tensor in the spherical coordinates of Eq.~\eqref{eq:TPE} requires a rotation by the spiral angle $\psi$, $K_{rr}=K_\parallel\cos^2\psi+K_{\perp r}\sin^2\psi$, $K_{\phi\phi}=K_\parallel\sin^2\psi+K_{\perp r}\cos^2\psi$, $K_{r\phi}=(K_\parallel-K_{\perp r})\sin\psi\cos\psi$, and $K_{\theta\theta}=K_{\perp\theta}$, with $\tan\psi=B_\phi/B_r$.
This work adopts a typical empirical expression for $K_\parallel$ from Ref.~\cite{potgieterModulationGalacticProtons2014}:
\begin{equation}
K_\parallel = K_0\beta\left(\frac{B_0}{|\mathbf{B}|}\right)\left(\frac{R}{R_0}\right)^a\left[\frac{\left(\frac{R}{R_0}\right)^m+\left(\frac{R_k}{R_0}\right)^m}{1+\left(\frac{R_k}{R_0}\right)^m}\right]^\frac{b - a}{m}.
\label{eq:Kparallel}
\end{equation}
Here, $K_0$ is a normalization factor of the order of $10^{23}~\mathrm{cm^2\,s^{-1}}$, $\beta \equiv v/c$ denotes the particle speed relative to the speed of light,
$B_0$ is the reference magnetic field strength measured near Earth, and $R_0 = 1~\mathrm{GV}$ is the reference rigidity.
The expression constitutes a smoothed broken power-law, characterized by power-law indices $a$ and $b$ below and above the break rigidity $R_k=3~\mathrm{GV}$, respectively, and a smoothing parameter $m = 3$ that controls the sharpness of the transition.
The use of a transition rigidity and the choice of $R_k$ follow previous heliospheric modulation studies~\cite{potgieterModulationGalacticProtons2014,fiandriniNumericalModelingCosmic2021,jiangNewScenarioSolar2023}. Physically, $R_k$ characterizes the regime in which the particle Larmor radius becomes comparable to the correlation length of heliospheric magnetic turbulence. Because this correlation length varies appreciably with location and solar activity, we retain $R_k$ as a phenomenological scale of order a few GV. The transition separates two scattering regimes that determine the rigidity dependence of $K_\parallel$ and the interpretation of $a$ and $b$. Below $R_k$, resonant scattering by magnetic fluctuations with a power-law spectral index $\nu$ gives approximately $K_\parallel\propto\beta R^{2-\nu}$ ($a\simeq1/3$ for the Kolmogorov value $\nu=5/3$). Above $R_k$, where the Larmor radius exceeds the turbulence correlation length, the rigidity dependence steepens toward the $R^2$ asymptote~\cite{jokipiiPropagationCosmicRays1971,teufelAnalyticCalculationParallel2003,engelbrechtTheoryCosmicRay2022}.

The perpendicular diffusion has been realized to be anisotropic over the past two decades.
The perpendicular diffusion coefficient in the plane perpendicular to $\mathbf{B}$ is widely adopted as a fixed fraction of the parallel one, $K_{\perp r} = 0.02 K_{\parallel}$~\cite{potgieterModulationGalacticProtons2014}.
In contrast, since the first fast latitude scan of Ulysses it has been established that the polar perpendicular diffusion coefficient is enhanced relative to the in-plane one~\cite{heberCosmicRaysHigh2006, baloghGalacticAnomalousCosmic2008}.
The polar perpendicular coefficient is expressed as $K_{\perp\theta} = 0.02 K_{\parallel} f_{\perp\theta}$~\cite{potgieterModulationGalacticProtons2014}, where
\begin{equation}
f_{\perp\theta} = A^+ \mp A^- \tanh\left[-8(\left|90^\circ - \theta\right| \mp \theta_F)\right],
\label{eq:f_theta}
\end{equation}
with $A^\pm = (d \pm 1)/2$, $\theta_F = 35^\circ$, and an enhancement factor of $d = 3$ towards the poles.
The ratio $0.02$ and the polar enhancement factor $d=3$ are prescribed a priori rather than fitted in this work. They are to some extent degenerate, because no data outside the ecliptic are used here, and they should therefore be regarded as adopted assumptions rather than as independently constrained quantities.

Defining the unit vector along the magnetic field as $\mathbf{e}^B$, the contribution from the antisymmetric component of the diffusion tensor can be reformulated as a cross product.
Using index notation, where $i$, $j$, and $k$ label the three spatial directions, repeated indices are summed, and $\epsilon_{ijk}$ is the Levi-Civita symbol, and noting that contraction with the symmetric part of the derivative vanishes, the expression simplifies as follows:
\begin{equation}
\begin{aligned}
\nabla\cdot(\mathbf{K}_A\cdot\nabla f) &= \partial_i\left(-\epsilon_{ijk}K_A\mathbf{e}^B_k\partial_jf\right) \\
&=-\epsilon_{ijk}\partial_i(K_A\mathbf{e}^B_k)\partial_jf - \epsilon_{ijk}K_A\mathbf{e}_k^B\partial_i\partial_j f \\
&= \nabla\times(K_A\mathbf{e}^B)\cdot\nabla f.
\end{aligned}
\end{equation}
This result indicates that antisymmetric diffusion produces an effect analogous to a drift velocity, defined by $\mathbf{V}_d \equiv \nabla \times (K_A \mathbf{e}^B)$.
Here, $k$ in $\epsilon_{ijk}$ is a summed spatial index and is unrelated to the subscript $k$ denoting the break rigidity $R_k$.

Under the weak-scattering assumption, the particle drift velocity is given as
\begin{equation}
  \mathbf{V}_d = \frac{\beta R}{3}\nabla\times\left(\frac{\mathbf{B}}{B^2}\right).
  \label{eq:V_d}
\end{equation}
Substituting the magnetic field from Equation~\eqref{eq:B_field} into Equation~\eqref{eq:V_d} yields the detailed expressions for the drift velocity in two distinct regions.
For particles located away from the HCS, the drift velocity arises from the gradual gradient and curvature of the HMF~\cite{jokipiiEffectsParticleDrifts1977}:
\begin{equation}
\begin{aligned}
  \mathbf{V}_d^\mathrm{GC} =& \frac{2 \beta R r}{3 B_0 r_0^2 (1 + \gamma^2)^2}\left[1 - 2\mathrm{H}(\theta-\theta_\mathrm{cs})\right] \\
 &\times \left(-\frac{\gamma}{\tan\theta}\hate_r+(2+\gamma^2)\gamma\hate_\theta+\frac{\gamma^2}{\tan\theta}\hate_\phi\right),
\end{aligned}
\label{eq:V_d_gc}
\end{equation}
where $\gamma= r\Omega_\odot\sin\theta/\mathrm{V_{SW}}$.
Near the HCS, the sharp magnetic field transition contributes an additional drift term.
Assuming a locally flat current sheet, the magnitude of this term was derived and numerically computed~\cite{burgerDriftTheoryCharged1985}.
This numerical solution can be parameterized as~\cite{burgerInclusionWavyNeutral1987, peiThreedimensionalWavyHeliospheric2012}
\begin{equation}
\begin{aligned}
   V_d^\mathrm{HCS} =& c \beta\mathrm{H}(2r_g - d) \\
   & \times \left[ 0.457 - 0.412 \left(\frac{d}{r_g}\right) + 0.0915\left(\frac{d}{r_g}\right)^2\right],
\end{aligned}
\label{eq:V_d_hcs_mag}
\end{equation}
where $d$ denotes the distance to the current sheet, $r_g$ is the Larmor radius, and the Heaviside function restricts this term to within twice the Larmor radius.
Calculating the distance $d$ requires scanning the entire HCS.
To achieve this efficiently, we employ an iterative one-dimensional optimization procedure in the $r$ and $\phi$ directions to locate the minimum distance.
This algorithm typically converges in fewer than thirty iterations.

The HCS drift direction is perpendicular to the magnetic field and parallel to the current sheet.
Considering the HCS shape described in Equation~\eqref{eq:theta_cs}, the HCS drift component is given by:
\begin{equation}
  \begin{aligned}
    \mathbf{V}_d^\mathrm{HCS} =& V_d^\mathrm{HCS} \times \left[
      \frac{r\sin\theta_\mathrm{cs}\Omega_\odot}{V_n}\hate_r + \frac{\mathrm{V_{SW}}}{V_n}\hate_\phi \right.
       \\
      &\left. - \frac{\tan\alpha\sin\theta_\mathrm{cs}\cos\phi_0(\mathrm{V^2_{SW}} + r^2\sin^2\theta_\mathrm{cs}\Omega_\odot^2)}{\mathrm{V_{SW}}V_n}\hate_\theta \right],
  \end{aligned}
  \label{eq:V_d_hcs}
\end{equation}
where $\phi_0 = \phi + \Omega_\odot r/\mathrm{V_{SW}} - \Omega_\odot(t-t_0)$, and $V_n$ is a normalization factor ensuring the vector inside the square bracket is a unit vector.
Collectively, the gradient-curvature drift in Equation~\eqref{eq:V_d_gc} and the HCS drift in Equation~\eqref{eq:V_d_hcs} describe all drift effects within the HMF.

Altogether, the modulation model comprises five parameters: the reference magnetic field \(B_0\), the tilt angle \(\alpha\), and three parameters (\(K_0\), \(a\), and \(b\)) that describe the diffusion coefficient.
The parameter ranges, adopted from our previous work~\cite{jiangNewScenarioSolar2023}, are provided in Table~\ref{tab:modulation_parameter}.
The ranges for the spectral indices $a$ and $b$ have been updated to [0, 2] and [0, 3], respectively, to encompass their theoretically expected values derived from the power spectrum of the HMF~\cite{jokipiiPropagationCosmicRays1971}.

\begin{table}[h]
    \centering
    \begin{tabular}{lll}
        \hline
        \textbf{Parameter} & \textbf{Range} & \textbf{Units} \\
        \hline
        $B_0$             & [3, 8]         & $\mathrm{nT}$ \\
        $\alpha$          & [0, 75]       & $^\circ$ \\
        $K_0$             & [0.001, 1.5]   & $\mathrm{10^{23}\,cm^2\,s^{-1}}$ \\
        $a$               & [0, 2]     & - \\
        $b$               & [0, 3]     & - \\
        \hline
    \end{tabular}
    \caption{Allowed ranges for solar modulation parameters. These limits of $a$ and $b$ bracket the theoretical expectations of $a \approx 0.5$ and $b \approx 2$, derived from the magnetic-field power-spectrum analysis~\cite{jokipiiPropagationCosmicRays1971}.}
    \label{tab:modulation_parameter}
\end{table}

The three-dimensional Parker's equation, with the complex structure described above, is solved using a backward-in-time SDE method.
In this approach, pseudo-particles are traced backward from the Earth to the heliopause.
The evolution of a particle's position is governed by the SDE~\cite{yamadaStochasticViewSolar1998, zhangMarkovStochasticProcess1999, koppStochasticDifferentialEquation2012, kapplSOLARPROPChargesignDependent2016}:
\begin{equation}
\Delta\mathbf{r} = \left(\nabla\cdot\mathbf{K}_S - \mathbf{V}_{\mathrm{SW}} - \mathbf{V}_d^{\mathrm{GC}} - \mathbf{V}_d^{\mathrm{HCS}}\right)\Delta t + \symbf{\sigma}\cdot\Delta\mathbf{W},
\end{equation}
where $\Delta t$ is the time step, $\symbf{\sigma}$ is the matrix square root of the diffusion tensor $\mathbf{K}_S$, defined as $\symbf{\sigma} \equiv \mathrm{diag}\{\sqrt{K_{\perp r}}, \sqrt{K_{\perp\theta}}, \sqrt{K_\parallel}\}$ in a coordinate system aligned with the heliospheric magnetic field (HMF), and $\Delta\mathbf{W}$ is the increment of a Wiener process representing a three-dimensional standard normal random variable.
The corresponding evolution of a particle's momentum is then described by
\begin{equation}
\Delta p = -\frac{p}{3r^2}\frac{\partial r^2\mathrm{V_{SW}}}{\partial r}\Delta t.
\end{equation}

Tens of thousands of particles with varying energies were simulated using the SDE described above.
The transition matrix of the spectrum was then computed statistically to characterize the modulation process.
For this implementation, we have independently developed HELPROP\footnote{\url{https://github.com/linsujie/Helprop}}, a software package specifically tailored to the present model. It is not derived from any existing code, but it belongs to the same family of time-backward SDE solvers of Parker's transport equation as SolarProp~\cite{kapplSOLARPROPChargesignDependent2016}, HelMod~\cite{boschiniPropagationCosmicRays2018}, Geliosphere~\cite{geliosphereSolanik2023}, and SDEMMA~\cite{sdemmaSong2025}.

\begin{figure}
    \centering
    \includegraphics[width=0.45\textwidth]{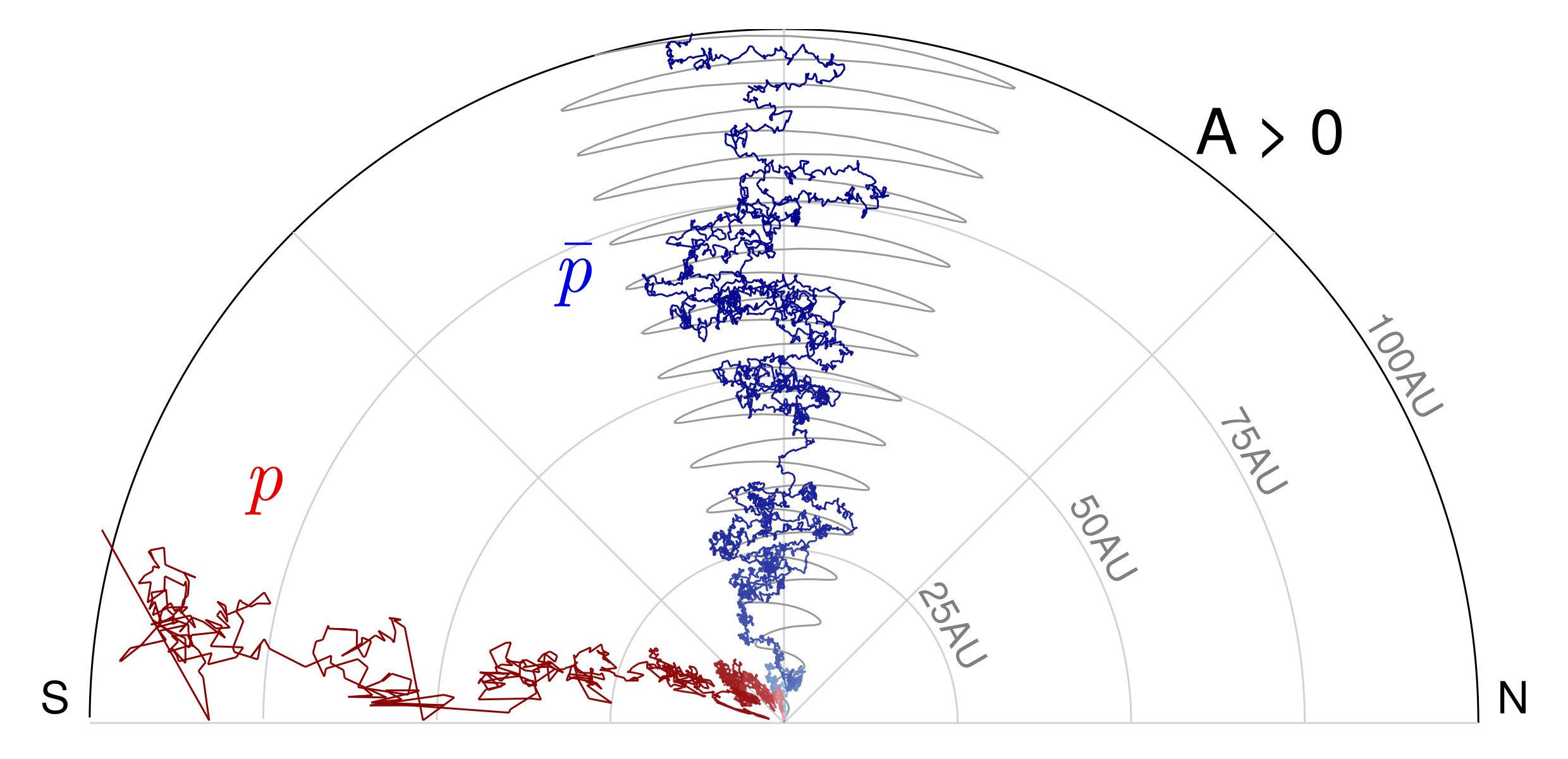}
    \caption{Representative particle trajectories simulated with HELPROP for A > 0 polarity. The red and blue curves correspond to the proton and antiproton tracks, respectively. The gray wavy line denotes the HCS.}
    \label{fig:particle_trace}
\end{figure}

The physically realistic HCS drift effect implemented in HELPROP plays a critical role in the particle transport within the heliosphere.
As an illustration, Figure~\ref{fig:particle_trace} shows the trajectories of two sample particles with rigidities near $1\,\mathrm{GV}$ propagating through the HMF during an A > 0 magnetic polarity.
Under this polarity, antiprotons experience a strong inward HCS drift, causing most antiprotons to enter the heliosphere primarily along the HCS.
In contrast, protons are strongly drifted outward by the HCS drift, while the gradient‑curvature drift in the inner heliosphere tends to transport them toward the HCS.
Consequently, only those protons that cross the polar regions without being diverted toward the HCS can avoid the strong outward HCS drift and ultimately reach Earth at 1 AU.

\section{Model Fitting}

This study aims to perform a global fit to determine the favoured parameter spaces of both the Galactic cosmic-ray propagation and solar modulation models.
However, since a single GALPROP calculation requires several minutes and a HELPROP calculation can take tens of minutes, Monte Carlo sampling of the parameter space becomes computationally prohibitive.
To address this challenge, we employ machine learning to train surrogate models that emulate these two processes, as detailed in Section~\ref{sec:machine_learning}.

Using these efficient emulators, we subsequently fit the AMS-02 and Voyager data.
The configuration for this fitting procedure is described in Section~\ref{sec:fitting_configuration}.

\subsection{Machine Learning Model}
\label{sec:machine_learning}

The propagation of CRs in the Galaxy and their subsequent heliospheric modulation are fundamentally linear processes.
This linearity implies that the influence of models such as GALPROP and HELPROP on the final CR spectrum can be effectively described by a linear transformation matrix, with each distinct set of model parameters corresponding to a unique matrix.
To capture this relationship, we developed a dedicated artificial neural network (ANN) architecture capable of learning the mapping from parameters to matrix.
This network is applied separately to each physical process, generating the precise transformation matrix for any given parameter set of the respective model, thereby explicitly preserving the linearity inherent in the propagation problem.
We refer to this framework, which uses two separate networks to emulate GALPROP and HELPROP respectively, as "PropMat".

\begin{figure*}[t]
    % \centering
    \includegraphics[width=0.8\textwidth]{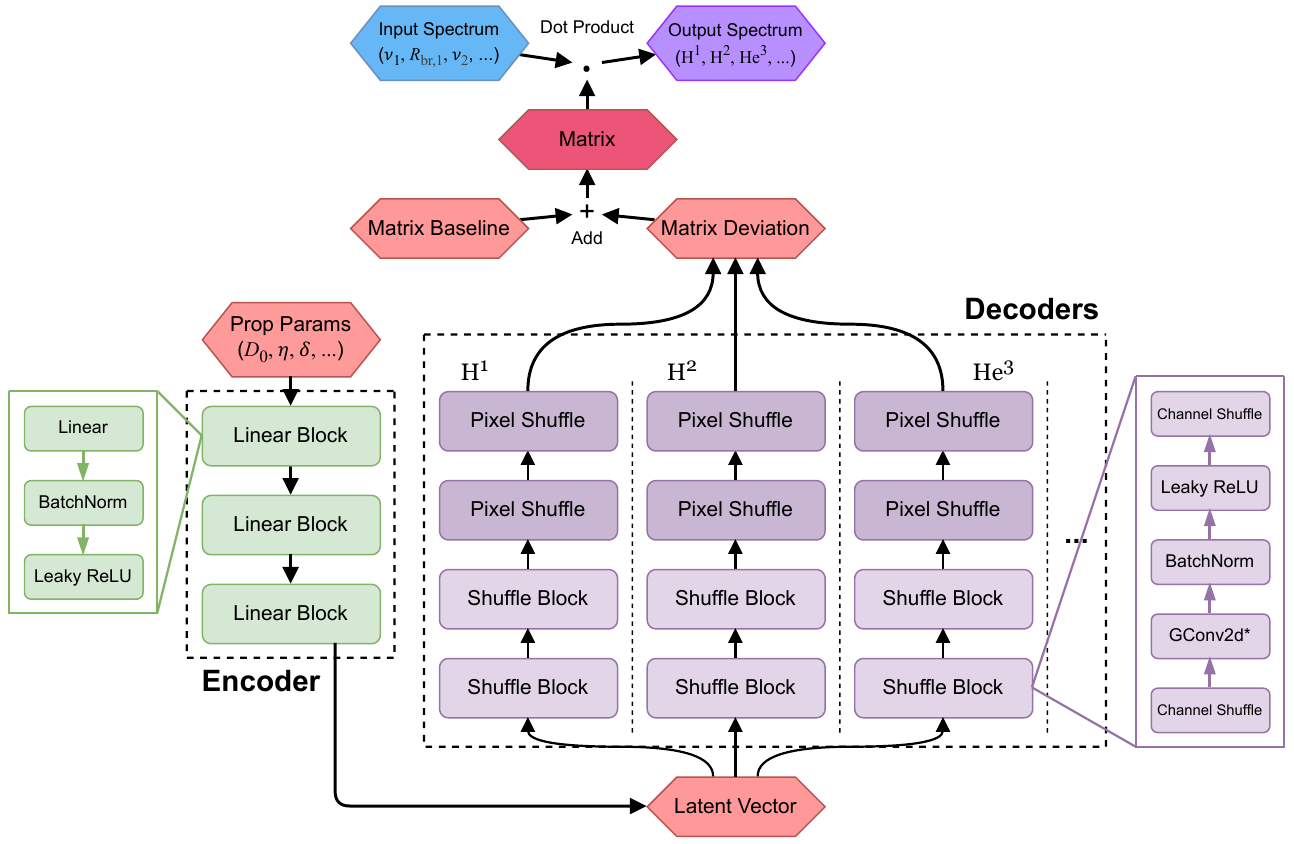}
    \caption{Deep learning architecture for spectral data processing in space physics, featuring propagation matrix (PropMat) generation and analysis across spatial domains (Galaxy and Heliosphere) using an encoder-decoder framework. The GConv2d operator depicted in the figure denotes a two-dimensional grouped convolution layer.}
    \label{fig:PropMat}
\end{figure*}

The architecture of PropMat is based on an encoder-decoder structure, as illustrated in Fig.~\ref{fig:PropMat}.
An initial Multi-Layer Perceptron (MLP) encoder maps the input model parameters into a latent representation, which extracts underlying features through nonlinear activation functions.
This latent vector is then decoded to produce the final transformation matrix.

A matrix generation via controlled multi-scale block operations is used to replace the conventional MLP decoder in PropMat.
The module employs PixelShuffle operations to enhance localized correlations across different spatial hierarchies.
In this framework, each matrix is dynamically split into independent groups at different hierarchical scales, with elements appear simultaneously in both higher- and lower-level groups.
The fully‑connected operations are then applied separately for all groups.
This process amplifies interactions among spatially adjacent positions, as they are linked through multiple connections across scales.
%The design enforces a hierarchical flow in which information propagates from coarse to fine blocks across stages—operating similarly to a multi‑scale block MLP.
Therefore, it inherently encourages spatial smoothness in the generated matrices while preserving computational efficiency through sparse group‑wise operations.

In order to train this ANN, we used the AdamWScheduleFree optimizer \cite{defazioRoadLessScheduled2024} for global optimization, followed by Stochastic Gradient Descent (SGD) \cite{bottouLargeScaleMachineLearning2010} for local fine-tuning.
AdamWScheduleFree is a schedule-free variant of AdamW that eliminates the need for manual learning rate scheduling.
Its interpolation-based update mechanism provides stable, competitive performance and enables rapid loss reduction during the initial training phase.
In contrast, SGD is employed for the final fine-tuning stage.
Its stochastic nature and simpler update rule help refine the model parameters and especially minimize prediction bias.

While the overall architecture of the two networks is similar, the loss functions for the GALPROP network and the HELPROP network are designed differently due to their distinct frameworks.

GALPROP employs discrete spatial and energy grids to model the evolution of particle densities during propagation. In this framework, the detailed transformation history of individual particles is not tracked.
Consequently, GALPROP outputs only a final LIS spectrum without providing the corresponding transformation matrix. The artificial neural network (ANN) for GALPROP must therefore learn to infer this implicit transformation matrix, $M$, from a set of LIS spectrum samples.
To enhance learning efficiency, we introduce a baseline matrix, $M_{\text{base}}$, obtained from multiple GALPROP runs with delta-function injections using a typical set of propagation parameters.
Assuming the true transformation matrix shares a similar structure with $M_{\text{base}}$, we decompose the predicted matrix as
\begin{equation}
   M_{\text{pred}}^{ij} = \exp\Bigl(\log M_{\text{base}}^{ij} + \log M_{\text{dev}}^{ij}\Bigr),
   \label{eq:M_base_dev}
\end{equation}
where the superscripts $i$ and $j$ label the energy bins of the injection and LIS spectra, respectively.
This allows the ANN to learn only the logarithmic deviation term, $M_{\text{dev}}$.
 To prevent the ill-conditioning of this term and mitigate over-fitting, we apply an L2 regularization penalty. Defined by the squared logarithmic deviations $\bigl[\log(M_{\text{dev}}^{ij})\bigr]^2$, this penalty favors $M_{\text{dev}}^{ij}$ values near unity.
The corresponding loss function is therefore defined as the mean-squared error (MSE) between the predicted and true spectra, combined with this regularization term.
Denoting the injection spectrum by $\psi_I$ and the output LIS spectrum by $\psi_O$, we have:
\begin{equation}
      \text{Loss}_{\text{G}}  = \sum_{j} \Bigl(\log\frac{\sum_i M_{\text{pred}}^{ij} \,  \psi_\mathrm{I}^{i}}{\psi_\mathrm{O}^{j}} \Bigr)^2 + \sum_{ij}\lambda \log \bigl(M_{\text{dev}}^{ij}\bigr)^{2},
     \label{eq:loss_GALPROP}
\end{equation}
where the MSE is evaluated on a logarithmic scale because the spectral values span several orders of magnitude.
The hyperparameter $\lambda$, which controls the regularization strength, is assigned with a relatively large value at first and decreased gradually.

\begin{figure}[h]
    \centering
    \includegraphics[width=0.4\textwidth]{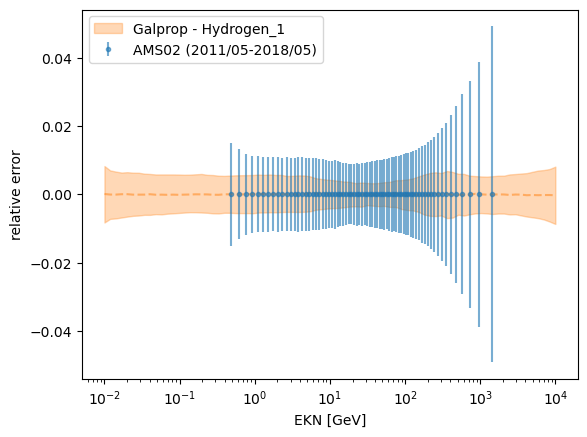}
    \caption{Prediction accuracy of proton's LIS, which is below the corresponding uncertainty of the AMS-02 experiment.}
    \label{fig:gp_accuracy}
\end{figure}

To generate the training data for this network, we employed a Markov Chain Monte Carlo (MCMC) algorithm\cite{foreman-mackeyEmceeMCMCHammer2013} within the parameter ranges specified in Table~\ref{tab:propagation_parameters}.
The procedure was designed to sample propagation parameters that yield acceptable fits to AMS-02 boron-to-carbon (B/C) ratio \cite{amscollaborationPrecisionMeasurementBoron2016} and to the Voyager B/C measurements \cite{cummingsGALACTICCOSMICRAYS2016}, adopting HELPROP for solar modulation. Approximately 25,000 samples were generated, covering a broad region of parameter space where $\chi^2/\mathrm{d.o.f} < 10$.
 These were split into a training set of 20,000 samples and a testing set of 5,000 samples.
As shown in Figure~\ref{fig:gp_accuracy}, the trained model achieves a relative error between predicted and true spectra within 1\% on the test set.
This level of accuracy is below the measurement uncertainties of the AMS-02 cosmic-ray data and will not introduce significant errors in subsequent analyses.

In contrast, HELPROP models heliospheric propagation by tracking a large ensemble of particles from Earth to the heliopause.
Each simulation therefore intrinsically provides the corresponding transformation matrix $M_\mathrm{O}$.
Although the matrix obtained from a finite particle sample is subject to statistical noise, we concentrate the sampling on a limited set of discrete energies to suppress fluctuations and then use interpolation to smoothly extend the matrix across the full energy range.
Hence the ANN for HELPROP can learn the transformation matrix directly, leading to a straightforward loss function defined as the MSE between the predicted and target matrices:
\begin{equation}
   \mathrm{Loss}_{\mathrm{H}} = \sum_{ij} \Bigl( \log \frac{M_\mathrm{pred}^{ij}}{M_\mathrm{O}^{ij}} \Bigr)^2,
\label{eq:loss_HELPROP}
\end{equation}
where the MSE is also evaluated on a logarithmic scale as in Equation~\eqref{eq:loss_GALPROP}.
Since the typical energy change during heliospheric modulation is on the order of the modulation potential (approximately $1\ \GeV$ for $|Z|=1$ particles), the energy-bin width in the transformation matrix must be considerably smaller than this scale.
We therefore adopt a linear binning scheme with a width of $0.05\ \GeV$ for all the results reported in this work.

\begin{figure}[h]
    \centering
    \includegraphics[width=0.4\textwidth]{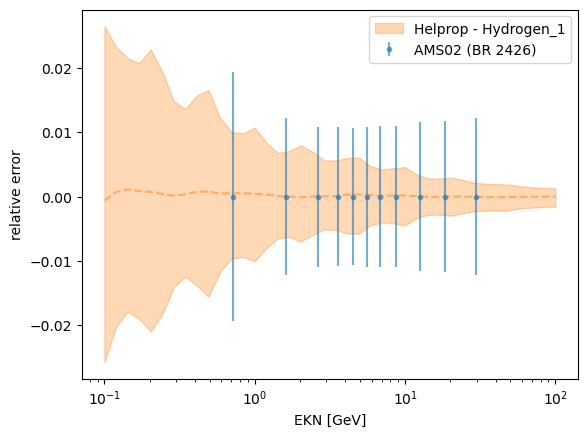}
    \caption{Prediction accuracy of the modulated proton spectrum, which is below the corresponding uncertainty of the AMS-02 experiment.}
    \label{fig:hp_accuracy}
\end{figure}

\begin{figure}[h]
    \centering
    \includegraphics[width=0.45\textwidth]{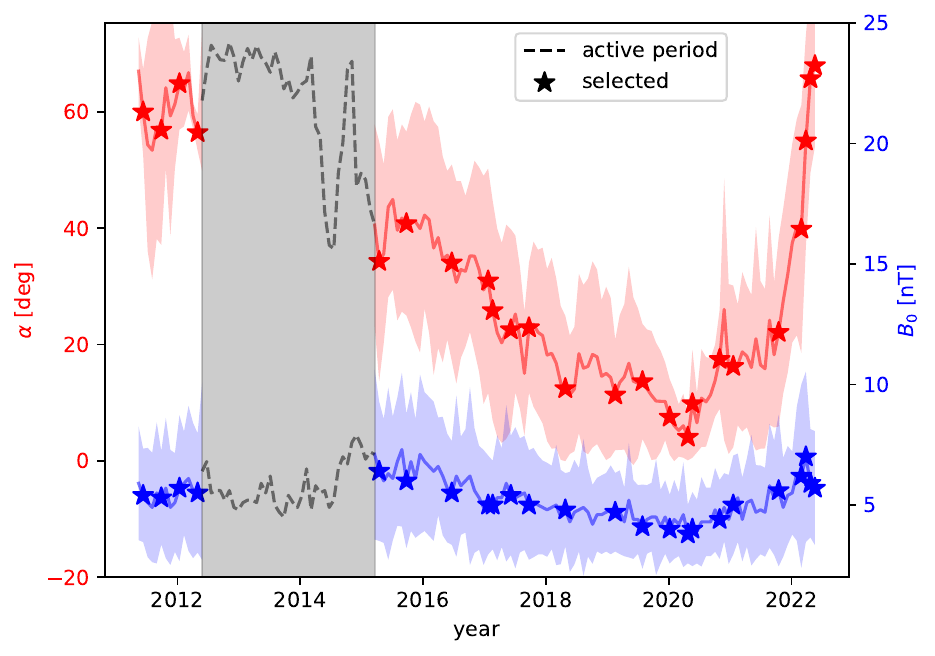}
    \caption{Magnetic field strength $B_0$ and tilt angle $\alpha$ as functions of Bartels rotation. The curves represent measurements from Ref. \cite{kingSolarWindSpatial2005,hoeksemaLargeScaleStructureHeliospheric1995}, with dashed segments corresponding to active periods and solid segments to solar minimum. The shaded bands indicate the observational uncertainties: the red band for $\alpha$ reflects the spread between the WSO line-of-sight ($L_\mathrm{av}$) and radial ($R_\mathrm{av}$) reconstructions, and the blue band for $B_0$ reflects the OMNI rotation-averaged measurement uncertainty. Stars mark the rotations selected for training during solar minimum.}
    \label{fig:B0_alpha}
\end{figure}

The training data for the HELPROP neural network were derived from the best-fit propagation parameters obtained during the GALPROP training phase.
Using an MCMC approach with a likelihood function analogous to that used for GALPROP, we fitted jointly to the Voyager LIS proton flux and the monthly AMS-02 proton flux~\cite{amscollaborationAntiprotonsElementaryParticles2025} by varying the injection and modulation parameters.
Twenty-four Bartels rotations were selected from the monthly data.
For each rotation, the magnetic field strength $B_0$ and the tilt angle $\alpha$ were constrained by Gaussian priors constructed from observational inputs (see Figure~\ref{fig:B0_alpha}), and the MCMC sampled the parameter space $(K_0, a, b, B_0, \alpha)$.
The Gaussian priors used here and in the subsequent global fit are constructed from the observational inputs shown in Figure~\ref{fig:B0_alpha}. For $\alpha$, the WSO potential-field reconstructions provide estimates of the average maximum latitudinal extent of the HCS from the classic line-of-sight model, $L_\mathrm{av}$, and the radial model, $R_\mathrm{av}$. We adopt $(L_\mathrm{av}+R_\mathrm{av})/2$ as the prior mean and $|L_\mathrm{av}-R_\mathrm{av}|$ as the prior standard deviation, thereby conservatively accounting for the systematic spread between the two reconstructions~\cite{hoeksemaLargeScaleStructureHeliospheric1995}. For $B_0$, the rotation-averaged field magnitude and its uncertainty from the OMNI measurements are used as the prior mean and standard deviation, respectively~\cite{kingSolarWindSpatial2005}.
The individual sample sets from all twenty-four rotations were merged into a combined training dataset.
In total, approximately 35,000 samples were generated and randomly split into a training set of 28,000 samples and a test set of 7,000 samples.
On the test set, the trained model for HELPROP attains a relative error between predicted and true spectra at the one-percent level over the energy range relevant to this work (see Figure~\ref{fig:hp_accuracy}).
This accuracy is sufficient to avoid introducing significant computational errors in the following calculations.

Although the training samples correspond to discrete points in the $B_0$–$\alpha$ plane, they are concentrated within a narrow band for the solar minimum period under study, thereby enabling the ANN to generalize effectively.
We have verified this generalization capability by evaluating the model on months excluded from the training set, confirming its consistent performance throughout the entire period of interest.
PropMat does not solve the transport equation; it interpolates the transformation matrix $M$ over the space of model parameters from a finite training set. Beyond the dense coverage of the parameters noted above, the accuracy of the interpolation requires $M$ to depend smoothly on them, which is a consequence of the diffusive character of the transport: diffusion damps sharp structure rather than preserving it, so perturbations of the transport coefficients cannot be amplified in the propagated and modulated spectrum.

%\begin{figure}[h]
%    \centering
%    \includegraphics[width=0.4\textwidth]{fig/gp_pred.png}
%    \caption{Prediction of LIS.}
%    \label{fig:hp_pred}
%\end{figure}

\subsection{Fitting Configuration}
\label{sec:fitting_configuration}

Utilizing the well-trained surrogate models, we conduct a global exploration of the parameter space using Markov Chain Monte Carlo (MCMC).
The fit incorporates monthly proton and antiproton flux data from AMS, the AMS boron-to-carbon (B/C) ratio, the local interstellar (LIS) proton spectrum and B/C ratio from Voyager, bounded by the propagation and modulation parameter ranges given in Table~\ref{tab:propagation_parameters} and Table~\ref{tab:modulation_parameter}.

To isolate charge-sign-dependent modulation effects, the analysis is confined to data from solar quiet periods and other non-reversal intervals between May 2011 and June 2022; the HMF polarity-reversal epoch (mid-2012 to early 2015) is excluded to avoid reversal-related complications~\cite{jiangNewScenarioSolar2023}.

Each Bartels rotation involves five free modulation parameters, yielding a total of 515 parameters over the 103 rotations of interest. Performing a full Monte Carlo sampling in this high-dimensional space is highly inefficient.
We therefore treat the modulation parameters as nuisance parameters.
The MCMC sampling is performed only over the several injection and propagation parameters, while the modulation parameters for each Bartels rotation are optimized separately within each likelihood calculation.
This reduces the effective sampling dimensionality to about ten, while still uniquely determining the best-fit modulation parameters for every rotation at each point in the reduced parameter space.

The observed CR fluxes at the TOA are shaped by source injection, Galactic propagation, and heliospheric modulation.
With the Galactic propagation constrained by the B/C ratio, the Voyager LIS proton spectrum can effectively anchor the source injection spectrum for primary protons.
The production of secondary antiprotons is then derived self-consistently from this framework.
Given these constraints, simultaneously fitting a unified modulation model to the measured TOA proton and antiproton fluxes provides a direct test for the charge-sign-dependent drift effects inherent to our modulation model.

\section{Results} \label{sec:results}

\begin{figure}[h]
\centering
\includegraphics[width=0.4\textwidth]{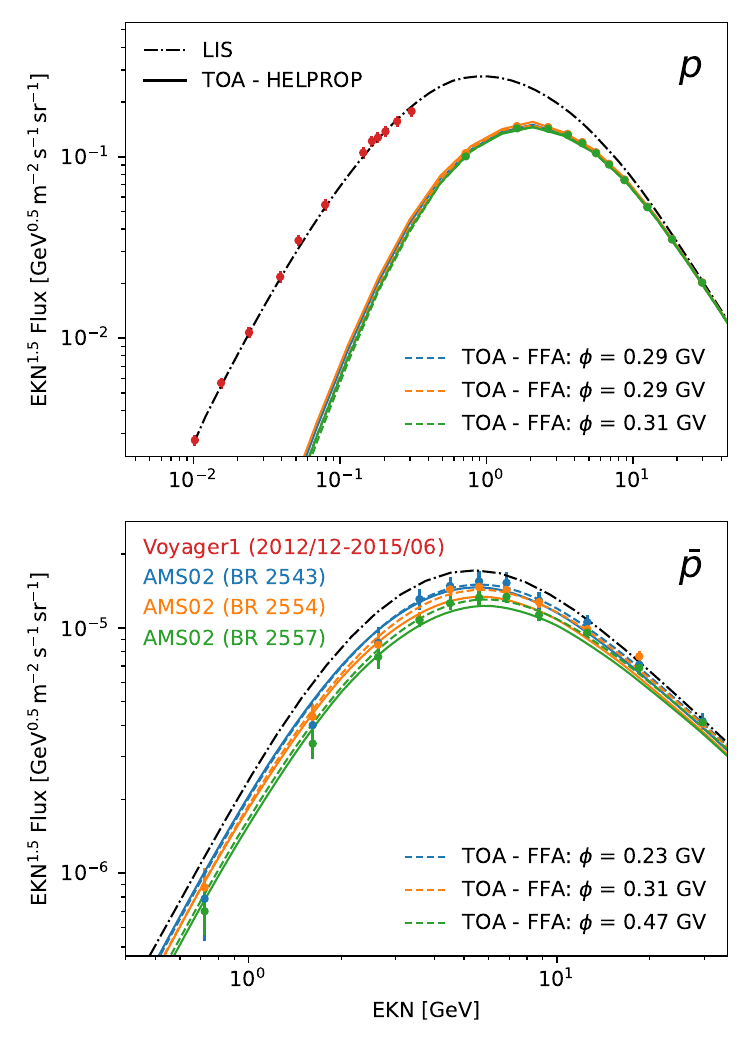}
     \caption{Best-fit proton (Top) and antiproton (Bottom) fluxes as functions of kinetic energy per nucleon (EKN). The black dash-dotted line corresponds to the LIS, while colored solid and dashed lines display spectra modulated using the HELPROP and FFA, respectively. Line colors match respective experimental measurements, with FFA modulation potentials $\phi$ indicated in the legend.}
\label{fig:best_fit_spec}
\end{figure}

The fit to the AMS B/C ratio, the Voyager proton LIS and B/C ratio, and the monthly AMS proton and antiproton fluxes yields a sensible fitting result.
HELPROP, which incorporates a realistic HCS drift effect, successfully describes the modulation of both protons and antiprotons within a unified framework.

Figure~\ref{fig:best_fit_spec} shows the best-fit local interstellar spectra for protons and antiprotons, along with the modulated spectra for several representative Bartels rotations, compared with Voyager and AMS-02 data.
For comparison, the modulation result obtained using the FFA is plotted with dashed lines.

We compare the local interstellar spectra derived here with those reported in Refs.~\cite{boschiniSolutionHeliosphericPropagation2017,boschiniInferenceLocalInterstellar2020,potgieterModulationGalacticProtons2014}. Over the rigidity range probed by AMS-02, our proton and antiproton spectra reproduce their rigidity dependence, while the normalizations differ by about $10$\,\%, which we take as the systematic uncertainty on the LIS input to our model.

The proton fluxes during Bartels rotations 2543, 2554, and 2557 exhibit close agreement, while the antiproton fluxes in these same rotations show significant discrepancies.
This flux disparity is primarily attributed to variations in the tilt angle $\alpha$.
During the studied period under positive magnetic polarity, as mentioned in Section~\ref{sec:modulation_model}, protons reach Earth primarily through the polar region and are thus insensitive to changes in the tilt angle.
In contrast, antiprotons detected at Earth are transported mainly via HCS drift.
A larger tilt angle extends the drift path, enhances particle losses during transport, and consequently lowers the observed flux.
Compared to the FFA, which requires distinct modulation potentials for protons and antiprotons to account for such differences, our model incorporates tilt angle information and successfully reproduces the observed variations for both particle species.

\begin{figure}[h]
\centering
\includegraphics[width=0.4\textwidth]{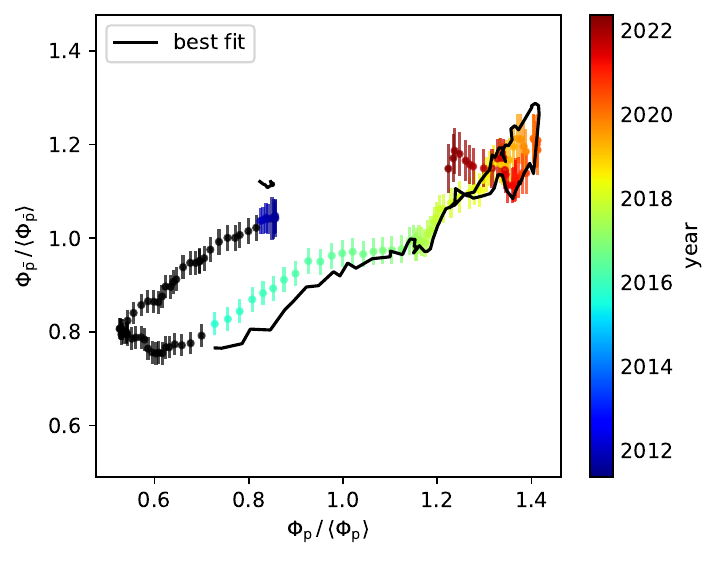}
\caption{
Correlation between antiproton and proton fluxes in the rigidity range of 1.00–2.97 GV, along with the corresponding best-fit result.
The horizontal and vertical axes show the proton and antiproton fluxes, respectively, each normalized to its time-averaged value over the full observation period.
 The data points represent the 13-Bartels-rotation moving averages of observed fluxes, normalized to their respective time-averaged values. The two lowest-rigidity bins, $1.00$--$1.92\ \mathrm{GV}$ and $1.92$--$2.97\ \mathrm{GV}$, are combined using weights proportional to their rigidity-bin widths. Colored points mark rotations included in the fit, with color indicating the year; black points mark active or polarity-transition periods, including rotations whose smoothing windows overlap the excluded HMF polarity-reversal epoch. The solid black line shows the best-fit model trajectory ordered by Bartels rotation and processed with the same smoothing and normalization as the data.
}
\label{fig:pbar_vs_p}
\end{figure}

This unified description of charge-dependent modulation can be further tested through the correlation between antiproton and proton fluxes over the full observation period, as shown in Fig.~\ref{fig:pbar_vs_p}.
Such charge-dependent modulation behavior is not limited to a few individual rotations but constitutes a systematic feature throughout the entire solar quiet period.
Figure~\ref{fig:pbar_vs_p} presents the correlation between antiproton and proton fluxes in the combined rigidity interval $1.00$--$2.97\ \mathrm{GV}$, which corresponds to a representative kinetic energy per nucleon of approximately $0.97\ \mathrm{GeV}$.
To mitigate the effects of short-term fluctuations, the data points are shown as 13-Bartels-rotation moving averages, normalized by their respective time-averaged flux values.
Our model consistently describes the flux variations of both charge signs across the entire observation interval, and reproduces the correlation between protons and antiprotons.
This further proves that the HELPROP model, based on a realistic three-dimensional heliospheric current sheet drift, can quantitatively and uniformly characterize charge-dependent modulation effects, outperforming the force-field approximation that requires independent modulation potentials for different charge signs.

The best-fit propagation and injection parameters are listed in Table~\ref{tab:best_fit_prop_inj}.

\begin{table*}[!hpbt]
    \centering
    \begin{displaymath}
    \begin{array}{*{9}{c}}
    \hline
    & \nu_1  & \nu_2 & R_\mathrm{br,1} & D_0 & \eta & \delta  & v_A & z_h \\
    & - & - & \mathrm{GV} & \mathrm{10^{28} cm^2\,s^{-1}} & -  & - & \mathrm{km\,s^{-1}} & \mathrm{kpc} \\
    \hline
    \textrm{Best-Fit} & 1.99 & 2.40 & 8.30 & 5.47 & -0.20 & 0.43 & 20.2 & 4.90\\
    \textrm{Med} & 1.99 & 2.39 & 8.25 & 5.44 & -0.19 & 0.43 & 20.3 & 4.85\\
    \sigma & 0.02 & 0.01 & 0.22 & 0.70 & 0.13 & 0.01 & 1.25 & 0.64\\
    \hline
    \end{array}
  \end{displaymath}
\caption{Best-fit values, posterior medians (Med), and standard deviations ($\sigma$) for the injection and propagation parameters.}
\label{tab:best_fit_prop_inj}
\end{table*}
All propagation parameters are well-constrained by the AMS and Voyager measurements of the B/C ratio together with a large set of antiproton and proton flux measurements.
The normalization of the diffusion coefficient, $D_0 = 5.44 \times 10^{28} \, \text{cm}^2 \, \text{s}^{-1}$, and the size of the propagation halo, $z_h = 4.85 \, \text{kpc}$, are determined within broad yet acceptable ranges.
The diffusion slope $\delta = 0.43$ is close to the expectation from the Kolmogorov spectrum ($\delta = 1/3$).
The Alfv\'{e}n speed $v_A = 20.3 \, \text{km} \, \text{s}^{-1}$ is consistent with typical interstellar medium values.

\begin{figure}[!hpbt]
\centering
\includegraphics[width=0.5\textwidth]{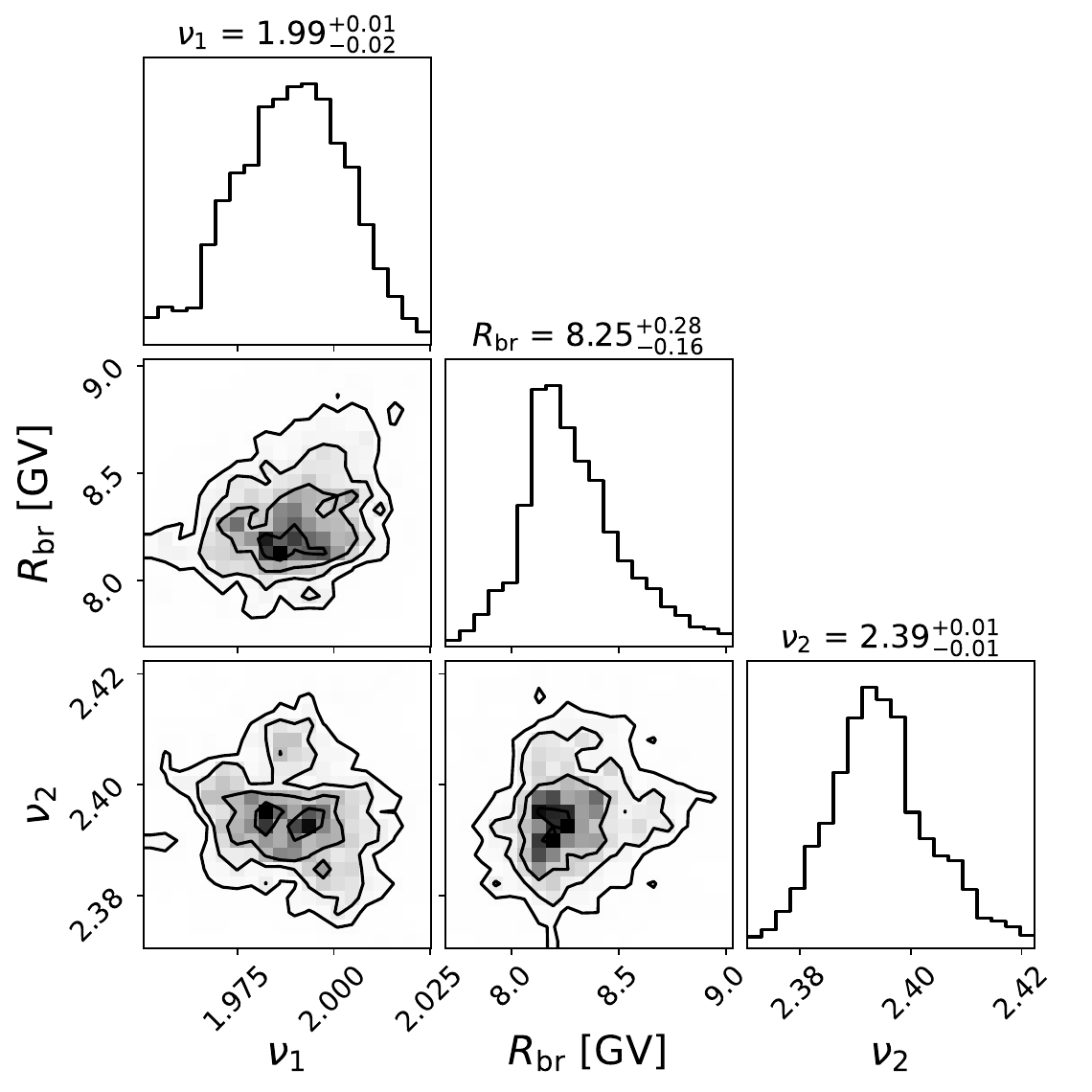}
\caption{One- and two-dimensional posterior distributions for the injection parameters $\nu_1$, $R_\mathrm{br,1}$, and $\nu_2$, as constrained by the AMS monthly data and the Voyager LIS. The diagonal panels show the marginalized one-dimensional posteriors; the labels report the medians and the corresponding intervals between the 16th and 84th percentiles. The off-diagonal panels show four contours enclosing approximately 11.8\%, 39.3\%, 67.5\%, and 86.5\% of the two-dimensional posterior probability.}
\label{fig:inject_fit}
\end{figure}

Only three of the five injection parameters are presented here.
This is because the AMS-02 monthly data used in the fitting are measured at energies below 30 GeV, while the second injection break is expected to occur above several hundred GeV.
Consequently, the parameters $R_{\mathrm{br,2}}$ and $\nu_3$ remain unconstrained by the data and were excluded from the fitting procedure.
Figure~\ref{fig:inject_fit} shows the one- and two-dimensional posterior distributions for the injection parameters $\nu_1$, $R_\mathrm{br,1}$, and $\nu_2$. The fit reveals a significant spectral break, with the index changing from $1.99$ to $2.39$ at a rigidity of approximately $8.25\ \mathrm{GV}$.
This break is effectively required by the spectral index of the Voyager LIS and the high-energy end of the AMS-02 data.

\begin{figure}[!hpbt]
\centering
\includegraphics[width=0.4\textwidth]{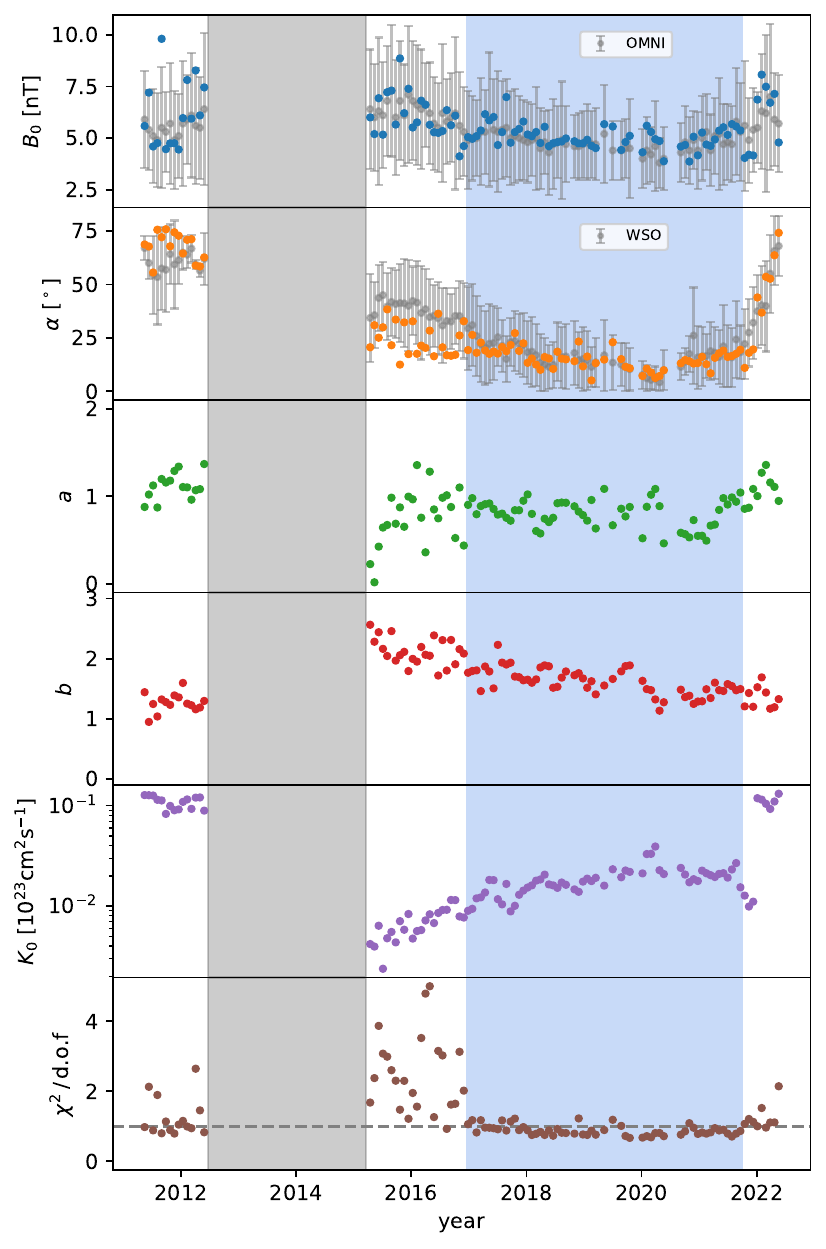}
\caption{Time variation of the HELPROP parameters. The observational estimates and uncertainties are overlaid on the $B_0$ and $\alpha$ panels; the uncertainty in $\alpha$ is defined as described in Sec.~\ref{sec:machine_learning}. The horizontal dashed line in the final panel marks the reference level of reduced $\chi^2 = 1$. The gray region marks the HMF polarity-reversal epoch excluded from the analysis, and the light blue region marks the BR 2502--2566 interval reliably described by the model.}
\label{fig:hp_evolution}
\end{figure}

Figure~\ref{fig:hp_evolution} displays the time variation of the modulation parameter.

The reduced $\chi^2$ for each Bartels rotation exhibits a significant correlation with the tilt angle $\alpha$.
Periods with large $\alpha$ are associated with high solar activity and a complex magnetic field, which increasingly disturbs the modulation process.
Consequently, the steady-state HELPROP model becomes gradually less accurate in these conditions, resulting in growing reduced $\chi^2$ values.
In extreme cases---for example, BR 2493 (April 2016)---the reduced $\chi^2$ can reach approximately 5.
Guided by Fig.~\ref{fig:hp_evolution}, we identify BR 2502--2566 (December 2016 to September 2021) as the interval over which the steady-state model provides a reliable description. Throughout this interval, the reduced $\chi^2$ remains below 1.3, the fitted tilt angle shows no systematic deviation from the observed values, and the reduced $\chi^2$ computed over the full interval is 0.87.

Outside this window, the model still captures the dominant trend in the data, but the description becomes progressively less accurate as the level and variability of solar activity increase. The excluded polarity-reversal epoch (mid-2012 to early 2015; gray region) coincides with the maximum of cycle~24.
Chronologically, the first interval to show noticeable deviations is BR 2426--2440 (May 2011 to June 2012), the only interval of negative HMF polarity in the sample. During this interval, the heliosphere was already approaching solar maximum, and the observed tilt angle was large ($54^\circ$--$67^\circ$).
The model nevertheless captures the overall modulation level: the reduced $\chi^2$ over the interval is approximately 1.2, the values for most rotations are below 1.15, and a few reach approximately 2.6. The fit quality is somewhat poorer than in BR 2502--2566: the fitted tilt angle exceeds the observed values by approximately $0.5\sigma$ on average, and, as shown in Fig.~\ref{fig:pbar_vs_p}, the antiproton fluxes around $1\,\mathrm{GeV}$ are somewhat overpredicted.
Even so, within BR 2426--2440 the fitted diffusion parameters remain well inside their prior ranges, $a\simeq0.9$--$1.4$, $b\simeq0.9$--$1.6$, and $K_0\simeq(0.8$--$1.3)\times10^{22}\,\mathrm{cm^2\,s^{-1}}$.

After the reversal (BR 2479--2501), the deviations become more pronounced: the reduced $\chi^2$ exceeds $\sim$3 for several rotations, and the fitted tilt angle is systematically lower than the observed values.
Toward the next solar maximum (BR 2567--2575), the observed tilt angle rises steeply. Although the reduced $\chi^2$ remains close to 1, the fits behave anomalously: the fitted tilt angle remains below the observed values during BR 2567--2569, and agreement with the observed tilt angle is restored only through an abrupt order-of-magnitude increase in $K_0$ at BR 2570, which interrupts the otherwise smooth evolution of $K_0$.
These deviations occur during high-activity epochs around the HMF polarity reversal or during episodes of rapidly changing tilt angle. We therefore restrict the following detailed discussion of the fitted modulation parameters to the reliably described interval.

Over BR 2502--2566, the reference diffusion coefficient at $1\,\mathrm{GV}$, $K_0$, increases steadily, reaching its largest values in 2020--2021 near the end of the solar minimum considered here.
This quantity has been estimated over a wide range in different studies, using 2D, 3D, and semi-analytical models, from $\sim\!10^{22}\,\mathrm{cm^2\,s^{-1}}$\cite{tomassettiEvidenceTimeLag2017,songNumericalStudySolar2021} up to $\sim\!10^{23}\,\mathrm{cm^2\,s^{-1}}$\cite{potgieterModulationGalacticProtons2014,kuhlenTimeDependentAMS02ElectronPositron2019}.
The largest values were reported for the unusually weak solar minimum of 2006--2009~\cite{potgieterModulationGalacticProtons2014}. The markedly weaker modulation during that period is naturally associated with larger diffusion coefficients, whereas analyses of the AMS-02 era tend to yield smaller normalizations~\cite{tomassettiEvidenceTimeLag2017}; our best-fit values are lower still. In addition to differing conventions for normalizing $K_\parallel$, cross-model comparisons are intrinsically limited because diffusion and drift jointly shape the modulated spectra and are partially degenerate in fits to observations, as explicitly demonstrated in the $K_0$--drift-scale plane~\cite{wangSolarModulationProton2022}. Models that treat drift differently---for example, by fitting a scaled drift coefficient rather than computing it at the weak-scattering value, as done here---therefore yield different effective diffusion normalizations. Quoted values of $K_0$ are thus comparable only at the order-of-magnitude level.

Within BR 2502--2566, the fitted rigidity indices remain well inside their prior ranges, with $a\simeq0.5$--$1.1$ and $b\simeq1.1$--$2.2$; hence neither is constrained by a prior boundary. These indices should be interpreted with care. They parameterize a phenomenological, path-averaged diffusion coefficient and effectively absorb spatial variations in turbulence along the propagation paths, the finite rigidity interval probed by the fit, and departures from the idealized assumptions of quasi-linear theory. They should therefore not be interpreted as direct measurements of the local turbulence spectral index $\nu$. With this caveat, the fitted values are consistent with the expected scattering regimes: $a$ lies somewhat above, but remains broadly compatible with, the range expected for resonant scattering in the inertial range of heliospheric turbulence ($a\simeq2-\nu$, i.e., $1/3$ for a Kolmogorov spectrum)~\cite{jokipiiPropagationCosmicRays1971,teufelAnalyticCalculationParallel2003,engelbrechtTheoryCosmicRay2022}, whereas $b\sim1$--$2$ is consistent with the transition toward the $R^2$ asymptote expected once the Larmor radius approaches the turbulence correlation length; a few rotations yield mildly steeper values~\cite{jokipiiPropagationCosmicRays1971,reichherzerTurbulencelevelDependenceCosmic2020}.

Both the HMF strength $B_0$ and the tilt angle $\alpha$ indicate a period of minimum solar activity around 2020.
Such conditions, characterized by weak magnetic fields and low turbulence, result in weaker confinement of low-energy particles.
This weaker confinement is reflected in the rise of the diffusion normalization $K_0$ toward its peak values in 2020--2021.
By contrast, the fitted low-rigidity index $a$ shows no monotonic trend over this interval. The weaker confinement of low-energy particles during solar minimum is therefore driven primarily by the increase in $K_0$, which directly scales $K_\parallel$ at all rigidities and thus increases the sub-GeV diffusion coefficient.

Thus, the solar minimum period is well described by HELPROP, and the behavior of the modulation parameter follows a reasonable trend within the model.

\section{Conclusion} \label{sec:conclusion}

We have developed a unified charge-dependent solar modulation model to explain the monthly proton and antiproton fluxes measured by AMS-02 during solar quiet periods.
In this model, the HELPROP, which incorporates a three-dimensional HCS drift effect, is employed to simulate drifts in a realistic manner, thereby accurately capturing the charge-dependent signatures in the modulation process.

Using the outputs of HELPROP and the galactic propagation code GALPROP, we separately constructed two surrogate models, called PropMat, based on a specially designed ANN architecture.
Trained on tens of thousands of simulated samples, each surrogate model reproduces the corresponding energy-transformation matrix of GALPROP or HELPROP for any parameter set within the observationally allowed region, with a computational cost of only $\mathcal{O}(\mathrm{ms})$.
The numerical error of these surrogate models is maintained at the one-percent level.
They can therefore be used to perform efficient global parameter searches for various study objectives.

In this study, we employed the HELPROP surrogate model to rapidly fit the monthly AMS-02 proton and antiproton fluxes, thereby validating the modulation model.
The LIS used in the modulation are derived from a GALPROP surrogate model and are further constrained by Voyager measurements.
The fitting results demonstrate that HELPROP can accurately describe both the proton and antiproton monthly fluxes within BR 2502--2566 while capturing their overall evolution elsewhere in the analyzed period, which encompasses both HMF polarity regimes; residual charge-sign-dependent systematics are concentrated in the high-activity epochs around the solar maximum and polarity reversal.
Moreover, it naturally accounts for charge-dependent effects in a unified manner.
The best-fit parameter values are physically reasonable, and their temporal evolution follows sensible trends.

This realistic drift-treatment approach should be applicable not only to protons and antiprotons, but also to electrons and positrons. In future work, we plan to apply the model to all these particle species using AMS-02 data during the solar minimum period. Furthermore, we will extend the model to periods with more complex heliospheric conditions.

This work is supported by the National Natural Science Foundation of China (NSFC) grants 12205388, 42150105, 12261141691, the Guangdong Provincial Key Laboratory of Advanced Particle Detection Technology (Grant No. 2024B1212010005), and the Guangdong Provincial Key Laboratory of Gamma-Gamma Collider and Its Comprehensive Applications (2024KSYS001). We thank the support from Fundamental Research Funds for the Central Universities, Sun Yat-sen University, No. 24qnpy123, and the Sun Yat-sen University Science Foundation.

\appendix
\section{Posterior Distributions of the Model Parameters}
\label{app:posteriors}

\begin{figure*}[!htbp]
    \centering
    \includegraphics[width=0.98\textwidth]{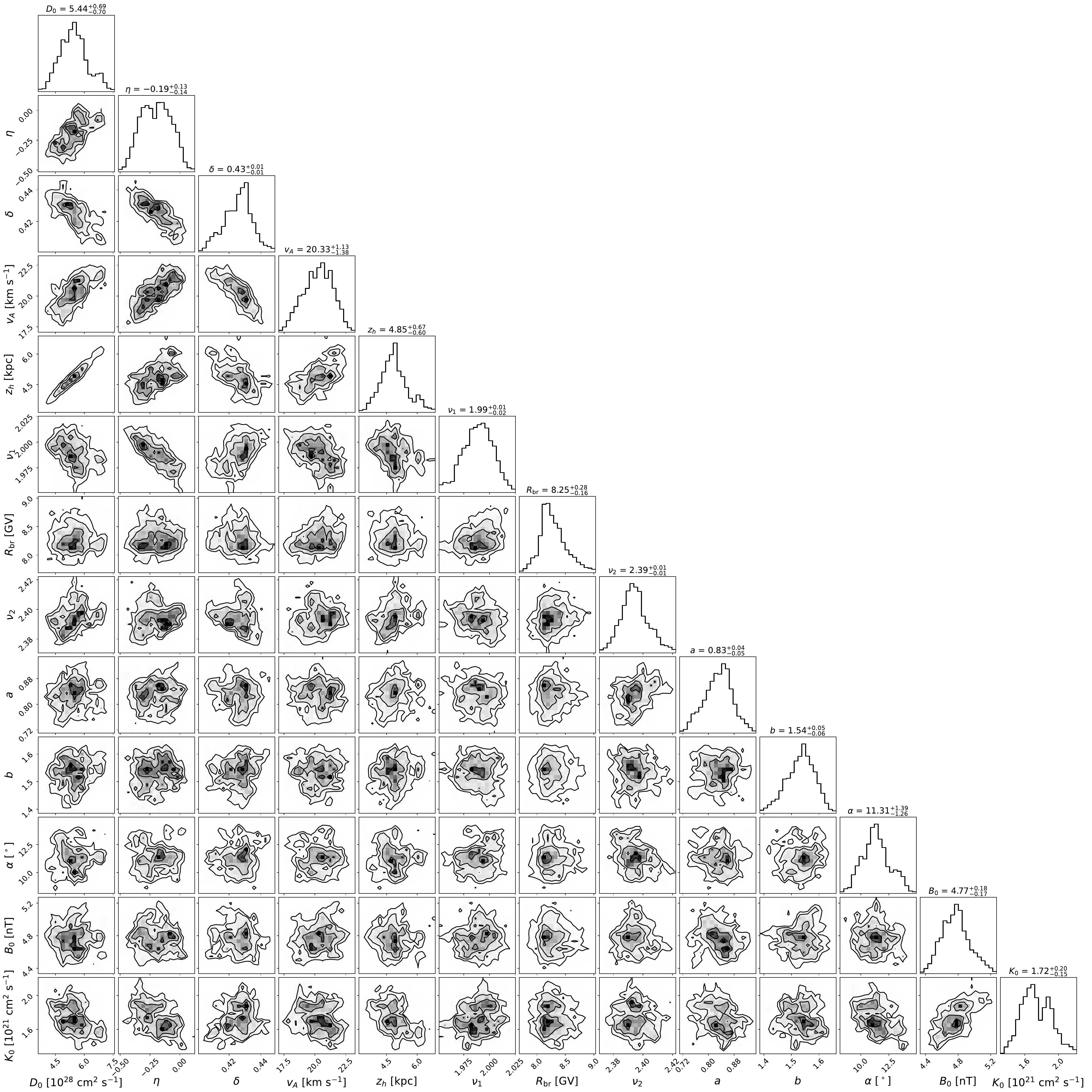}
     \caption{One- and two-dimensional posterior distributions from the global MCMC fit for the Galactic injection and propagation parameters and the solar modulation parameters at the representative Bartels rotation BR 2530. The diagonal panels show the marginalized one-dimensional posteriors; the labels report the medians and the corresponding intervals between the 16th and 84th percentiles. The off-diagonal panels show four contours enclosing approximately 11.8\%, 39.3\%, 67.5\%, and 86.5\% of the two-dimensional posterior probability.}
    \label{fig:triangle}
\end{figure*}

For a more detailed examination of the fitted parameters, Fig.~\ref{fig:triangle} shows the one- and two-dimensional posterior distributions of the Galactic injection and propagation parameters and the solar modulation parameters for a representative Bartels rotation.
We choose BR 2530 (January 2019) because it lies well within the solar quiet period and has a relatively low tilt angle, $\alpha \approx 14.4^\circ$. This value is inferred from the WSO observations as the mean of the line-of-sight and radial-model estimates, $(L_\mathrm{av}+R_\mathrm{av})/2$~\cite{hoeksemaLargeScaleStructureHeliospheric1995}.
Three-dimensional stochastic transport calculations show that cosmic rays can follow the wavy HCS effectively when the tilt angle is small, whereas at large tilt angles cross-field diffusion increasingly disrupts coherent drift along the HCS and reduces the sensitivity of the modulated intensity to $\alpha$~\cite{straussModellingHeliosphericCurrent2012,potgieterSolarModulationCosmic2013}.
BR 2530 therefore provides an illustrative low-activity case in which HCS drift is expected to remain relevant.

\bibliographystyle{utphys}
\bibliography{charge_depend_modulation}
\end{document}